\documentclass[reprint,
 amsmath,amssymb,
 aps,pre
]{revtex4-2}
\usepackage{graphicx}
\usepackage{dcolumn}
\usepackage{bm}

\usepackage{xcolor}
\def\Ri{R_i}
\def\et{\epsilon_{\theta}}

\newcommand{\dc}{\,$^{\circ}\mathrm{C}$}

\begin{document}

\preprint{APS/123-QED}

\title{The crack pattern of indented cohesive granular media}
\author{Marie-Julie Dalbe}
\author{Pierre Jodlowski}
\author{Nicolas Vandenberghe}%
\affiliation{Aix Marseille Univ, CNRS, Centrale Med, IRPHE, Marseille, France}

\date{\today}

\begin{abstract}
Cohesive granular materials, such as wet sand, retain their shape before yielding under stress, exhibiting a solid-like behavior. As the loading increases, the material typically flows. However, cohesive materials can also develop cracks, similar to those observed in brittle materials. This study investigates the formation of cracks during the indentation of a granular pile. A solid cylindrical indenter is pushed quasi-statically into the wet granular medium, causing radial cracks to appear. These cracks resemble mode I cracks observed in brittle elastic solid materials. We characterize the crack pattern through direct observation and three-dimensional X-ray microtomography. Notably, we establish a correlation between the dilation of the material during shearing and the appearance of cracks. 
\end{abstract}

\maketitle


\section{Introduction }

Pushing an intruder into a granular medium is a classic problem used to test our understanding of granular flows. It is also a loading configuration commonly encountered in human activities, such as the compaction of powders during manufacturing~\cite{CJM2013} or material testing~\cite{BLV1980,MHB2021} or in natural situations, like the locomotion of terrestrial animals~\cite{SKG2015} and the formation of craters by celestial impacts~\cite{Katsuragi2016}. This configuration has been extensively studied to assess the penetration force in relation to the properties of the medium~\cite{KFL2018}. The quasi-static penetration of a blunt indenter, like a flat-ended cylinder, is accompanied by a plastic flow that has been characterized using X-ray microtomography in a dry granular medium~\cite{MSC2006}. The grains are pushed downwards under the indenter and sideways, resulting in a divergent axisymmetric flow. The flow leads to a local change of density, with the domain underneath the indenter experiencing compaction while on the side of the indenter, dilation is observed. 

Adding cohesion forces between the individual spheres changes the overall picture. Cohesive granular materials~\cite{MN2006,Her2005}, such as wet sands, are able to sustain shear stress even in the absence of a confining pressure~\cite{RER2006}. When the stress is progressively increased, above a critical stress, the medium yields. The resulting flow leads to the rearrangement of the network of contacts and cohesive bonds break and form. In addition to this shear response, cohesive granular materials are also able to sustain a tensile load. Remarkably, in certain stress configurations, when a critical stress is reached, cracks develop in the granular medium. They differ from the shear bands that are observed in many instances of complex fluid flows~\cite{DFM2016} and notably in cohesive granular materials~\cite{MNP2021}. In the present work, we discuss tensile cracks that are similar to the mode I cracks of solid brittle material~\cite{Broberg1999}. In a solid material, the formation of cracks is generally interpreted as the nucleation and extension of a surface of discontinuity in a stressed sample. Crack extension is made (energetically) possible by the release of stored potential elastic energy. In cohesive granular material, cracks have been studied in specifically designed loading cells~\cite{PC1997,SSB2008} and in bending tests~\cite{RB2018}. The critical stress at which the cracks develop has been linked to the inter-grain cohesive force but such studies have been limited to configurations for which the plastic flow is limited or well controlled. Therefore the state of the material at the onset of fracture can easily be described with reference to the initial, unstressed state, similarly to the approach taken in solid materials. However cracks also develop in situations where the sample experiences large plastic deformation. Cracks have  been observed in various instances of complex fluid flow~\cite{RMJ2013,BBD2022} and cohesive granular material~\cite{SSX2024}. The circumstances under which a cohesive granular material will develop a tensile crack rather than flow when experiencing loading remains elusive. 

The goal of the present work is to describe the appearance of cracks and to study their pattern when a rigid cylinder is quasi-statically pushed into a wet granular medium for which cohesion is caused by liquid bridges between the grains. Radial cracks develop on the side of the cylinder. Such radial cracks pattern have been observed in impacted cohesive powders~\cite{CCM1970} and in brittle materials such as glass, but they are not ubiquitous as conical or ring cracks are often observed~\cite{LOL2009,RJR2021}. Here we use direct pattern observation and X-ray microtomography to document the morphology of cracks and the deformation field in the sample at the onset of fracture.

The paper is organized as follows. We first discuss the experimental apparatus and the preparation of wet granular medium samples used for the indentation experiment. Then, we present experimental results obtained through direct observation of the experiment. Our primary focus is on the pattern of radial cracks and its variation with indenter size, grain size and liquid content. We then describe and analyze X-ray microtomography scans conducted during the indentation experiments. These scans are used to compute the deformation field, which reveals that the region of apparition of cracks is characterized by large dilation rather than large orthoradial strain. Finally we discuss the experimental results in relation with the material properties.

\section{Experimental setup and phenomenology }

\begin{figure}
    \centering
    \includegraphics[]{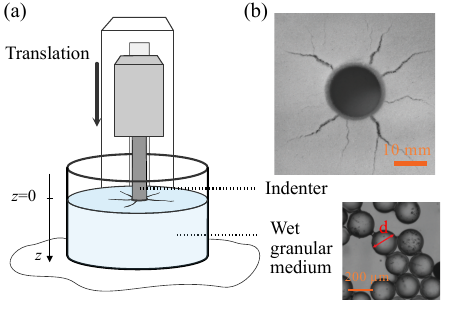}
    \caption{(a) Sketch of the experiment. A cylindrical indenter is slowly pushed into a cohesive granular material made of glass beads of controlled diameter wet with silicon oil.  The inset (bottom right) shows some grains under a microscope. (b) The typical crack pattern exhibits radial cracks (picture taken from above at the end of the experiment, after removal of the indenter).}
    \label{fig:exp}
\end{figure}

We use a cohesive granular medium made of glass beads with a controlled diameter. Different media with different bead diameters $d$ between $55$~$\mu$m and $1.15$~mm were used for the experiments (see Supplemental Materials Table I for the details of the material used). For these sizes, Van der Waals interactions are negligible. The glass density is $\rho_g = 2500$~kg~m$^{-3}$. Before use, the grains are soaked in water and soap inside an ultra-sonic bath at 55\dc. They are then heavily rinsed with de-ionized water, and dried in an oven at 80\dc ~for at least 10 hours. 

We  mix the grains with a silicone oil of viscosity $2 \times 10^{-3}$~Pa~s (Roth 2277.1) using a spatula. The oil density is $\rho_{l} = 870$~kg~m$^{-3}$ and its surface tension is $\gamma \approx 20$~mN~m$^{-1}$.
The amount of oil added to the grains is measured by the weight ratio defined as $w = \text{oil mass} / \text{grain mass}$. Most of our experiments are conducted with $w = 0.5\%$ or $w = 8\%$. For $w=0.5\%$, the saturation (volume of liquid divided by volume of voids) is $1.9 \%$ and the medium is in the pendular regime where we expect individual liquid bridges at the contact points between the grains. The medium with $w=8 \%$ is in the funicular regime, and liquid filled voids are expected \cite{MN2006}. 

The experimental set-up is shown in Fig.~\ref{fig:exp}. A mass $m$ of cohesive grains is poured inside a cylindrical container of diameter $D_c = 10$~cm and compressed using a flat lid so the grains reach a height $h_g\approx 8$~cm. The resulting volume fraction of the pile is $\phi_0 = m / [ \pi h_g (D_c / 2)^2 \rho_g (1+w) ]$. Throughout all the experiments, we find a reproducible volume fraction $\phi_0 = 0.57\pm 0.02$. We therefore consider that our initial state is the same in all the experiments. This is in agreement with previous work~\cite{FY1998}. However the granular medium is heterogeneous, with local densities that can vary. Other preparation methods should be used to limit this phenomenon~\cite{SKG2015}. 

We indent the material using a cylinder of radius $\Ri$ pushing with a linear stage at velocity $1$~mm~s$^{-1}$. The indenter penetrates down to a depth of about $20$~mm. As the indenter penetrates the granular pile, we observe the formation of cracks in the granular medium expanding radially from the indenter. At the end of the experiment, the indenter is moved upwards and removed to take a picture of the final pattern (see Fig.~\ref{fig:exp}(b) and SM Fig.1 of the Supplemental Materials). We did not observe a significant change of the pattern during the removal of the indenter. We used different indentation speeds between $0.1$~mm~s$^{-1}$ and $4$~mm~s$^{-1}$, as well as step by step motion and did not observe a change of pattern. We thus consider that in the regime considered here, this experiment is independent of the rate of loading. We also neglect the inertia forces of the grains.   

The depth of penetration of the indenter is denoted $z_i$ and $z_i = 0$ corresponds to the first contact of the indenter with the surface of the granular pile.  The radial cracks appear at a critical indentation depth  $z_i = z_c$. Cracks are not observed with dry grains. On the final pattern, we count the number of cracks $N$. After their first observation, as $z_i$ increases, cracks extend radially and widen. Cracks present some variability, some being wider and more visible than others. They may divide (showing branches), but we do not observe the formation of new cracks near the indenter. The number $N$ used to compute the wavelength $\Lambda = 2 \pi R_i / N$ is the number of cracks before branching, close to the indenter. We observe radial cracks for all the tested indenters (radii from $\Ri = 2$~mm to $\Ri = 15$~mm). We observe radial cracks for the grains tested in the range $d=55$~$\mu$m up to $d\approx500$~$\mu$m. 

For grains of diameter $d=1.15$~mm, we do not observe any crack. The Bond number $Bo = \rho_s g d^2 / \gamma$ measures the ratio of the characteristic stress due to gravity  $\rho g d$ and  cohesion stress $\gamma / d$. As $d$ increases, the Bond number increases and thus the effect of cohesion when compared to gravity decreases. When $Bo$ reaches a value of order unity, capillarity is comparable to gravity, and is not sufficiently intense to maintain the grains together and the material behaves more like a dry material. 

\begin{figure}
    \centering
    \includegraphics{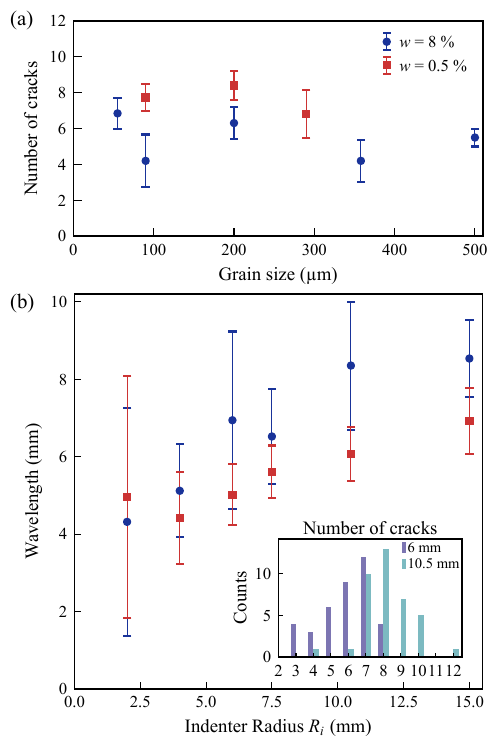}
   \caption{(a) The number of radial cracks for different grain sizes $d$ for an indenter radius $R_i = 7.5$~mm. The data points are the average values and the half-length of the error bars is the standard deviation of the distribution of cracks numbers. We do not observe a clear trend in the change of the number of cracks with grain size $d$. (b) The wavelength $\Lambda = 2 \pi R_i / N$ of the crack pattern for different indenter radii. Each data points correspond to the average and the error bars to the standard deviation of at least 10 experiments combining different values of grain size $d$. The wavelength increases for small $R_i$ but the growth rate seems to be weaker for larger $R_i$ values. The data points correspond to two liquid contents ($w = 0.5 \%$: red squares and $w = 8 \%$: blue dots). The inset shows two typical distributions of the number of cracks for $w = 8\%$ and two different $R_i$}
    \label{fig:NCracks}
\end{figure}

For the same experimental conditions, corresponding to given values of indenter radius $R_i$, grain size $d$ and liquid content $w$, with the same cell geometry, the experiment is repeated several times (at least 10 times). The granular medium is carefully mixed and compacted between experiments. The outcome of the experiments, \textit{i.e.} the counted number of cracks, is statistically distributed, as expected from fragmentation experiments. From the set of experiments, we compute an average and a standard deviation. In the inset of Fig.~\ref{fig:NCracks}(b), we show the distribution of the number of cracks for two different values of $\Ri$ for a liquid content $w = 8\%$. The experiments are repeated for distinct values of the grain size $d$,  distinct values of the indenter radius and distinct values of the liquid content. When $d$ is varied, we do not observe a clear change of the number of cracks (see Fig.~\ref{fig:NCracks}(a)).  
In  Fig.~\ref{fig:NCracks}(b), we show the variation of the wavelength when $R_i$ is varied. The dataset contains measurement with different $d$ values.  
The number of cracks $N$ increases with $\Ri$, and the wavelength $\Lambda = 2 \pi \Ri / N$ increases at small $R_i$ with a slope $\Delta \Lambda / \Delta R_i$ that decreases as $R_i$ increases. We also observe a difference between the data sets at $w = 0.5 \%$ and $w = 8 \%$: at higher liquid contents a smaller number of cracks is observed. 

\begin{figure}
    \centering
    \includegraphics{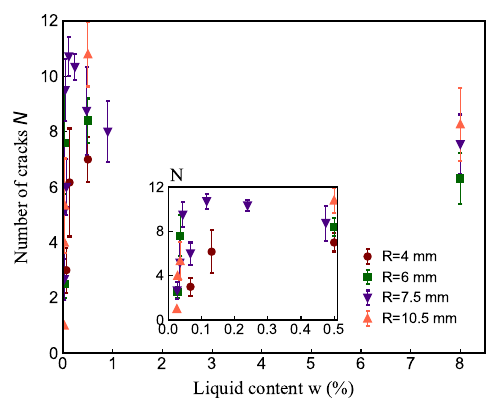}
    \caption{The number of cracks $N$ for different liquid contents $w$ for $d=200$~µm and different $R_i$. The inset shows the same data for $w<0.5 \%$. The number of cracks increases sharply for $w<0.2 \%$. The larger data sets are obtained for $w = 0.5 \%$ (pendular regime) and $w = 8 \%$ (funicular regime). A slight diminution of the number of cracks is observed between the pendular and funicular regimes.} 
    \label{fig:Nvsw}
\end{figure}

If changing $d$ is a way to change the intensity of the cohesion, changing the liquid content $w$ also affects the cohesion but in a more complex fashion~\cite{SSB2008,FGH2005,RB2018}. In Fig.~\ref{fig:Nvsw} we plot $N$ versus $w$. We observe a sharp increase for low values of $w$ (typically $w < 0.1 \%$). For the range of $w$ shown in the inset of Fig.~\ref{fig:Nvsw}, the granular medium is in the pendular regime. In previous studies, in this regime, all the tested mechanical properties (tensile stress, critical acceleration, cohesive stress, differential pressure in a shearing experiment) were almost independent of $w$~\cite{SSB2008,FGH2005,RB2018}. This was explained by the Laplace pressure and the shape of the bridges : in the pendular regime, each bridge is independent. The Laplace pressure decreases when the size of the bridge increases while the area of the bridge increases. The cohesive force is thus approximately constant~\cite{SSB2008}. In this regime, the only dependence on $w$ is for very small liquid content, where the number of bridges per bead increases sharply from 0 to 6. Fournier et al. found an experimental value $w_b\approx0.04\%$~\cite{FGH2005} which would be in agreement with our experimental results. However, they find that this critical value depends on the bead roughness, which we have not measured.

In the funicular regime, beads are no longer connected by individual bridges, but by fluid clusters. This changes drastically the pressure distribution, and as a consequence most mechanical properties are affected. The cohesive stress increases with liquid content~\cite{RB2018}, the differential pressure in a shearing experiment decreases with $w$~\cite{FGH2005}. However, the tensile strength and critical acceleration for fluidization remain constant with $w$~\cite{SSB2008}. In this regime the mechanical properties are governed by the Laplace pressure in the clusters, which reaches a critical value, and the projected area does not depend on the size of the cluster. 
In our case, we do not find a strong change in the mean number of cracks between the pendular regime ($w=0.5\%$) and the funicular regime ($w=8\%$) though we note a small but robust decrease in the number of cracks from the pendular to the funicular regime (see Fig~\ref{fig:NCracks}).

\section{Microtomography measurements}
X-ray microtomography is used to measure the precise topology of the medium during crack extension and the displacement field within the granular medium. We indent the material in steps of 0.5~mm, and make a 3D scan at each step. Each scan has a resolution of 65~$\mu$m/voxel. These experiments were performed with grains of diameter $d=200$~$\mu$m. We add a small amount of markers, steel beads of diameter 250~$\mu$m (0.15\% in volume), to allow three-dimensional tracking from which we compute the displacement fields with a custom made program. Four experiments were performed with different $\Ri$ (7.5 and 10~mm) and different liquid content ($w=0.5\%$ and $8\%$) (Table \ref{tab:tomo}).  

\begin{table}[tb]
    \centering
    \begin{tabular}{p{3cm}|c|c|c|c}
        Liquid content $w$  & \multicolumn{2}{c|}{$w=0.5\%$} & \multicolumn{2}{c}{$w=8\%$}  \\
        Indenter radius $\Ri$ (mm)& $7.5$ & $10$ & $7.5$ & $10$ \\
         \hline
         Indenter depth at onset of fracture $z_c$~(mm) & $3$  & $2.7$  & $3$  & $3$\\
         Mean crack depth at onset $z_m$ (mm) & $0.5\pm 0.2$ &  $1.2\pm 0.4$ & $0.7\pm 0.6$ & $1.3\pm 0.6$ \\
         Mean crack radius at onset $r_m$ (mm) & $11\pm 2$ &  $17\pm 2$ & $13\pm 2$ & $18\pm 2$ \\              
    \end{tabular}
    \caption{Crack characteristic lengths for the different experiments in the CT Scanner. $z_c$ is the indenter position at the scan where cracks are first observed. $z_m$ and $r_m$ are the mean depth and radius of the cracks when they first appear.  }
    \label{tab:tomo}
\end{table}

\subsection{Crack morphology}
\begin{figure}[htb]
    \centering
    \includegraphics{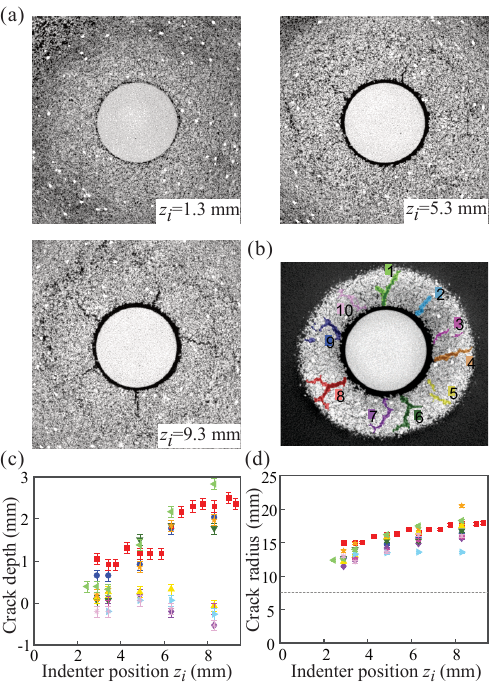}
    \caption{Crack morphology obtained from microtomography scans of  an experiment with $d=200$~µm, $\Ri = 7.5$~mm and $w=8\%$. (a) Slices at $z = 0.59$~mm depth for three different penetration depths $z_i$. Cracks are visible at $z_i = 5.3$~mm and $9.3$~mm. (b) A slice in the $(x,y)$ plane showing the pattern of cracks at a depth $z = - 1.6$~mm for an indenter depth $z_i = 9.3$~mm. When cracks are first observed, they appear with a finite depth (c) and radial extension (d). For better visibility, we plot the values of crack depth and crack radius for all the $z_i$ only for crack 8.}
    \label{fig:CrackLength}
\end{figure}

Direct visual observation at the surface of the sample during indentation does not permit an unambiguous determination of the appearance of cracks. To gain some insight into the crack morphology, including their appearance,  we examine the slices in the $(x,y)$ planes obtained from microtomography. Typical slices at a depth $z = 0.59$~mm is shown in Fig.~\ref{fig:CrackLength} (a). Cracks appear at a given indentation depth $z_i = z_c$ of about 3~mm. When cracks are first observed in a scan, they have a well defined depth  (maximal value of $z$ at which the crack is observed) and radius defined as the distance between the crack tip and the center of the indenter (Fig.~\ref{fig:CrackLength} and table \ref{tab:tomo}). When the cracks presents multiple branches, the largest radius is used. As indentation depth increases, most cracks extend, but at rates that can be very different from crack to crack. For example in the experiment of Fig.~\ref{fig:CrackLength}, the increase rate for the depth vary from 0 (cracks 2, 3, 5, 10, which are the fainter cracks, do not expand in depth) to around 0.4 for the other cracks. The increase rate for radius goes from 0.1 (crack 2) to around 1. We also notice that the fainter cracks do expand in radius at the same rate as others. 

\subsection{Displacement fields and dilation}
\begin{figure}[ht]
    \centering
    \includegraphics{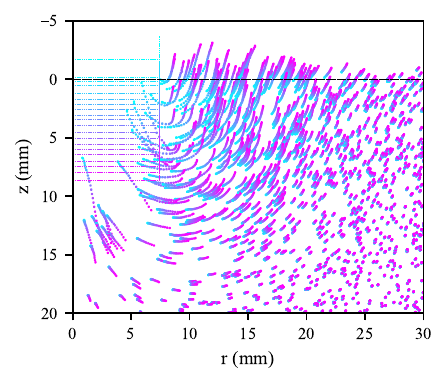}
    \caption{Successive positions of tracers in the $(r, z)$ plane. Experiment with $w=8\%$, $\Ri = 7.5$~mm and $d=200$~$\mu$m. The colors correspond to the indenter depth $z_i$: blue is at the beginning of the experiment, pink at the end. The indenter positions are shown on the upper left. 500 tracers out of around 40000 are shown. Tracers from different azimuthal planes are used as axisymmetry is assumed. }  
    \label{fig:traj}
\end{figure}

The displacement field for a typical experiment is shown in Fig.~\ref{fig:traj}. It is constructed by tracking the motion of 40000 markers in a volume of about 400~cm$^3$ with a precision of 0.02~mm. Despite the presence of cracks, we observe a radial symmetry of the displacement field with a high degree of precision and thus we plot the trajectories in the $(r,z)$ plane combining markers for all azimuthal angles. At frame $k$ after contact, each marker, indexed by $n$, has coordinates $(r_n (k), z_n (k))$. $R$ and $Z$ refer to the Lagrangian coordinates, \textit{i.e.} to the initial position in the pile. 

Three domains can be identified in Fig.~\ref{fig:traj}.  Markers initially close to the surface below the indenter (\textit{i.e.} small $Z/R_i$ and $R < R_i$) move with the indenter in the $z$-direction as it penetrates the granular pile. Markers initially located further below the surface ($Z/R_i \gtrsim 1$ or $R \gtrsim R_i$) are pushed sideways. On the side of the indenter the beads have an upward trajectory. The surface rises: at the end of the experiment, some beads are at $z<0$, with $z=0$ corresponding to the original surface position. Finally markers that are located further sideway, typically $R > 3 R_i$, do not move significantly. 
This overall picture is  consistent with the scenario of the formation of a cone of grains underneath the indenter that, once formed, moves together with the indenter, like a solid body~\cite{KFL2018}. Outside this cone, grains in the vicinity of the indenter (\textit{i.e.} for $r_n(0)$ typically less than  about $3 R_i$) present a typical diverging motion. In addition we also note that the displacement field between two scans keeps the same shape as it is translated when $z_i$ increases. 

\begin{figure*}[htb]
    \centering
    \includegraphics{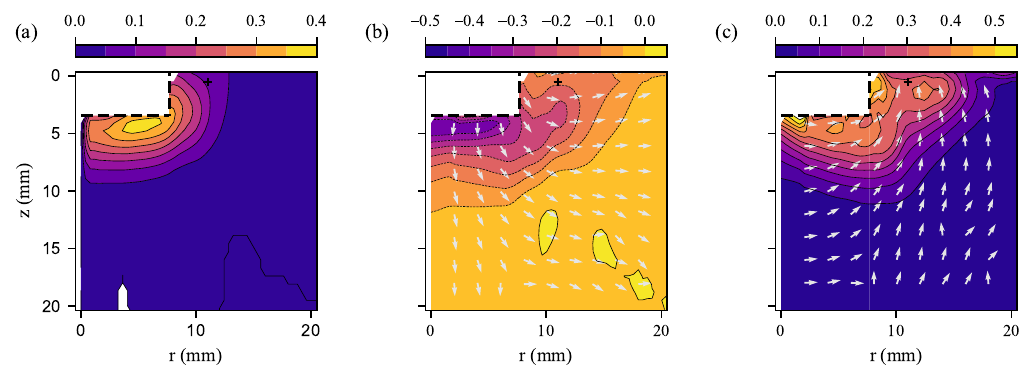}
    \caption{The field of eigenvalues of the Green-Lagrange strain tensor. (a) The eigenvalue associated with the ortho-radial direction exhibits a positive ortho-radial strain underneath the indenter, but its location does not correspond to the locus of apparition of cracks. (b,c) The two other eigenvalues are shown with the eigenvectors. The average location of the crack tip at nucleation is shown by the black cross. Experiment with $w=0.5\%$, $R_i=7.5$~mm, $d=200~\mu$m for $z_i\approx z_c$.  }
    \label{fig:vf}
\end{figure*}

For each tracer identified by its index $n$, the displacements for scan $k$ are $u_{r}(k) = r_n(k) - r_n(0)$ and $u_{z}(k) = z_n(k) - z_n(0)$. 
To gain a better description of the strain experienced by the material, we compute the derivatives of the displacement field and the Jacobian $\nabla_X \underline{u}$ of the displacement field (see SI for the details of the computation, including averaging over all azimuthal angles, filtering and fitting procedure).  The eigenvalues of the Green Lagrange strain tensor $\mathbb{E} = (1/2) ( \mathbb{F}^T \mathbb{F} - \mathbb{I})$, where $\mathbb{F} = \mathbb{I} + \nabla_X \underline{u}$, are related to the change of lengths in the deformed medium~\cite{Holzapfel2000}. This analysis is particularly relevant to the discussion of the length of a capillary bridge. Consider a segment between two neighboring bead centers. The center-to-center vector is initially $\overrightarrow{\delta}_0$ where $\lVert \overrightarrow{\delta}_0 \lVert = d$ for touching spheres. After indentation, when the indenter is located in $z=z_i$, the distance between the centers is $\lVert \overrightarrow{\delta} \lVert$ and $\lVert \overrightarrow{\delta} \lVert^2 - \lVert \overrightarrow{\delta}_0 \lVert^2 = \overrightarrow{\delta}_0 \cdot \mathbb{E}  \overrightarrow{\delta}_0$. Therefore the eigenvalues of $\mathbb{E}$ are directly related to the change of length of the center-to-center vector. With respect to radial cracks, the eigenvalue $\lambda_2$ associated with the ortho-radial direction, with an eigenvector $\overrightarrow{e_{\theta}}$, is the most relevant. The field of $\lambda_2$ exhibits a maximum underneath the indenter and this location does not correspond to the locus of crack nucleation (Fig.~\ref{fig:vf}). A distinctive feature of the flow of a wet granular material is the ability to break capillary bridges -- as would be expected for the development of cracks -- but also to form new bridges when two spheres are brought into contact by the displacement field. The formation of new bridges will be particularly favored in area where the density of beads is high. In the domain with large orthoradial positive strain, underneath the indenter, we also observe a strong negative (or compressive) strain associated with the vertical motion. Therefore, we expect the formation of new bridges in this area possibly preventing the formation of cracks. Hence, because of the strong plasticity of the wet granular medium, it is necessary to consider the local density of the medium. 

\begin{figure}[htb!]
    \centering
    \includegraphics{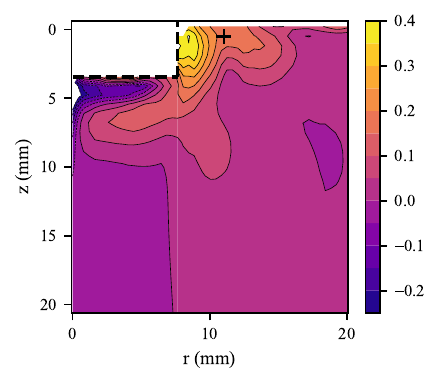}
   \caption{(top) Dilation field $\Delta V / V_0 = J - 1 $ at $z_i=3$~mm, with $w = 0.5\%$, $R_i = 7.5$~mm and $d = 200$~µm, corresponding to the frame at which cracks are first observed. The area underneath the indenter is compacted while the area on the side of the indenter is dilated. The area with a positive dilation on the side of the indenter is also the domain where cracks appear. The average location of the crack tip at nucleation is shown by the black cross.}
    \label{fig:dilation}
\end{figure}

It has long been known that granular flows are accompanied by changes of the local density. An initially compact medium will dilates as it flows while an initially loose medium will compact. The dilation field measures the change of density as the material is deformed. The measurement of the displacement field can be used to compute the volume ratio $J = \mathrm{det} \,\mathbb{F}$ after indentation. The dilation or normalized change of volume  $ \Delta V / V_0 = J - 1$ is plotted in Fig.~\ref{fig:dilation} for a specific experiment at the scan where cracks are first observed. Remarkably the dilation field shows a maximum at the location of crack initiation, shown by a black cross. This observation suggests that cracks in a cohesive granular medium is associated with the dilatancy of the medium \textit{i.e.} the expansion of the medium as it is sheared. Cracks appear in the area with the low volume fraction. It should be noted that the smallest volume fraction inferred from the displacement field in Fig.~\ref{fig:dilation} is  $\phi_{low} \approx 0.41$. This value is clearly low but it is of the same order as the volume fraction of the stable pile obtained by carefully sifting wet grains through a wide grid (typical sieve size of 3 mm). The high value of the volume fraction obtained near the symmetry axis ($R=0$) is about $\phi_{high} \approx 0.68$ but it should be taken cautiously because the density of markers goes to zero at the center of the cell. The error in computing the strain $\epsilon_{\theta} = u_r / R$ is thus amplified by low values of $R$. 

Additional features of the dilation field are also presented in supplementary materials. As the indenter penetrates into the granular medium the dilation in the vicinity of the indenter increases in amplitude linearly with the indenter penetration $z_i$ but the characteristic length of the dilation field remains constant, of the order of $R_i$ (see Supplemental Materials Fig.~SM 2).

\section{Discussion}
\label{sec:model}

\subsection{Nucleation of cracks}
In Fig.~\ref{fig:dilation}, we show that the locus of apparition of cracks in an indented cohesive granular medium coincides with the domain of large dilation \textit{i.e.} weak density. The change of volume fraction resulting from the flow occurs with a rearrangement of the pile. A possible consequence of this rearrangement is the change of the typical distance between the spheres and thus an elongation of the liquid bridges between the grains. The behavior of a single liquid bridge of volume $V_{LB}$ between two solid spheres of diameter $d$ has been extensively studied~\cite{WAJ2000, Her2005}. As it is extended, the liquid bridge exerts a restoring force that decreases with the separation
\begin{equation} \label{eq:FLB}
F_{LB} = \frac{\pi \gamma d}{ 1 + 1.05  \hat{s} + 2.5 \hat{s}^2} 
\end{equation}
with $\hat{s} = s/ \ell$ with $\ell = (2V_{LB} / d)^{1/2}$ and $V_{LB}$ the capillary bridge volume. $s$ is the separation between the two spheres (see Fig.~\ref{fig:modWL} (c)). 
The liquid bridge breaks when the distance between the spheres reaches the critical separation $s_c$ given by~\cite{Her2005}
\begin{equation} \label{eq:sc}
\frac{s_c}{d} = \left( \frac{V_{LB}}{d^3} \right)^{1/3} + 0.2 \left( \frac{V_{LB}}{d^3} \right)^{2/3}
\end{equation} 
An estimate of $s_c / d$ can be made assuming the number of bridges per sphere is $\kappa \approx 6$ and all the liquid goes into individual liquid bridges and thus $V_{LB} = (\rho_g/\rho_l) w (2/\kappa) \pi d^3 / 6$. For $w=0.5 \%$, Eq.~\ref{eq:sc} yields $s_c / d \approx 0.14$. For $w = 8 \%$, the granular medium is in the funicular state and the assumptions that all the liquid goes into capillary bridges is clearly no longer valid. Individual capillary bridges coexist with voids entirely filled with liquid. A characteristic volume of the individual bridges can be computed by considering that, for $s = 0$, the critical volume of bridges at the limit of coalescence is $V_{LBc} \approx 0.0073 d^3$~\cite{WAJ2000, SSB2008}.  This gives $s_c / d \approx 0.20$. These values can be compared with our measurements of the dilation field. Assuming that in the initial state the spheres are in contact ($s=0$), for an isotropic deformation, without a change of the arrangement, the typical extension of bridges will be $s_c / d \approx (J-1)/3$. If the deformation is not isotropic, we could imagine it could reach a maximal value $s_c / d \approx (J-1)$ if all the deformation is in the $\theta$ direction. We can see in Fig.~\ref{fig:dilation} that we have the right order of magnitude ($s_c / d \approx 0.14$) , with $(J-1)$ between $0.4$ close to the indenter, to $0.2$ in the area of the crack tip. However, we would expect a larger value for $w=8\%$, which is not the case : we have roughly the same values for the dilation field (see Fig.~SM2 of the Supplemental Materials). This simple comparison of order of magnitudes should be taken cautiously as the assumption of an homogeneous isotropic deformation is clearly questionable. Indeed when sheared, cohesive granular materials tend to form clusters and thus to present a  heterogeneous structure. Clearly a more detailed exploration of the pile near the site of nucleation of cracks should be performed to get a better understanding of the dynamics before and during crack formation. 

\subsection{The pattern of cracks}

The second important observation of this study is the pattern of radial cracks that is selected at the early stage of the experiment. The number of cracks (or similarly the wavelength) evidently shows some dispersion in the experiments, but nonetheless, we measure a robust growth of the mean number of cracks when $R_i$ increases (see Fig.~\ref{fig:NCracks}). In several instances of crack pattern formation in solid materials, fracture mechanics has been used to analyze the spacing between cracks and relate it to material properties. Fracture mechanics is based on the interchange between the potential elastic energy stored in the medium and fracture energy associated with the creation of new surface and other dissipative effects occurring near the crack tip. A crack extends when the decrease of the potential energy stored in the sample caused by the crack moving forward is equal to the fracture energy necessary to break the material. Fracture mechanics has been very successful in the description of the motion of a single crack, but it has also been used when multiple cracks are interacting. For example, it has been used to predict the spacing between multiple parallel cracks in thermally quenched solids~\cite{BWB2010}, or the number of radial cracks in thin sheets~\cite{VVV2010, VVV2013}. These analyses have been conducted assuming an elastic material response and the crack spacing then depends on the ratio of the elastic modulus and the fracture energy. 

\begin{figure}[tbp!]
    \centering
    \includegraphics{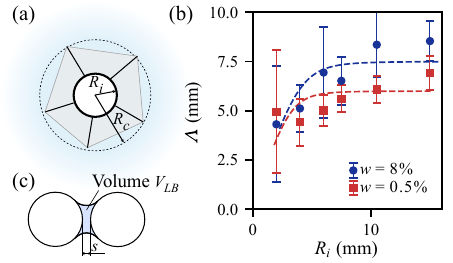}
   \caption{(a) Geometry of the model for crack extension. We focus on the area on the side of the indenter. The dashed circle shows the extent of cracks $R_c = \chi R_i$ and the grayed area shows the geometrical domain where the strain is released in the model. The blue area sketches the extent of the strained area. (b) Comparison of the model (eq.~\ref{eq:modelRes}) and the average wavelength obtained in the experiment. The experimental data includes the different values of $d$ (assuming that the wavelength is independent of the grain size). In the model, we adjust the length $\ell_c = \Gamma / (\sigma_0 \et)$ to fit the data. We obtain $\ell_c = 3.0 \pm 0.2$ ~mm for $w = 0.5 \%$ and $\ell_c = 3.75 \pm 0.25$ ~mm for $w = 8 \%$. (c) Sketch of the liquid bridge between two grains.}
    \label{fig:modWL}
\end{figure}

In the present case, we propose an \textit{ad hoc} model that is built on the competition between the potential energy and fracture energy. For cohesive granular materials, no clear and definitive description of the mechanical response is available. The model is based on two material parameters, the characteristic cohesion stress $\sigma_0$ associated with liquid bridges and the energy per unit surface required to extend a crack $\Gamma$. Note that $\Gamma$ is not the surface tension, as other dissipative effects may play a role in the crack opening process. An estimate of the potential energy that is stored in the strained material is based on the geometry shown in Fig.~\ref{fig:modWL}. The potential energy is associated with an ortho-radial strain $\et$. Because of this strain, capillary bridges are extended and we write the stored potential energy $U_{p} = \int \sigma_0 \et d\Omega$ to this state. We consider a  domain of height ($z$ direction) $h$. To keep the model simple we consider the case of a constant strain $\et$ from $r= \Ri$ up to $r = k \Ri>R_c$ where $k$ is a constant ($k \approx 3)$ as revealed by the dilation field measurements. $R_c = \chi R_i$ is the extent of the cracks. For $r > k R_i$, the strain is zero. In the uncracked state, the potential energy is $U_{p} = \pi \sigma_0 h \Ri^2 \et (k^2 - 1)$ because the strained volume is $\pi h R_i^2 (k^2 - 1)$. When cracks extend, the strain is released in their vicinity, leading to a decrease in the overall elastic energy. We compute the elastic energy of the cracked state by assuming that the stress is released in the polygon behind the lines connecting the crack tips (see Fig.~\ref{fig:modWL}). The elastic energy, for a pattern of $N$ regularly spaced cracks, extending to $R_c = \chi \Ri$, is then 
\begin{equation}
	U_{p} = \sigma_0 \et h \Ri^2 \left( \pi k^2 - \frac{N}{2} \chi^2 \sin \frac{2 \pi}{N} \right)
\end{equation}
The fracture energy is proportional to the crack surface 
\begin{equation}
	U_{f} = \Gamma 2 N h \Ri (\chi - 1)
\end{equation}
where $\Gamma$ is the energy needed to open a crack per unit crack surface

Minimization of the total energy $U_p + U_f$ with respect to the crack length $\chi$ yields
\begin{equation}
	\chi_N =  \frac{\Gamma}{\sigma_0 \et R_i} \left( \sin \frac{2\pi}{N} \right)^{-1}
\end{equation}
To determine the optimal number of cracks of the pattern, minimization with respect to $N$ for $\chi = \chi_N$ is performed, similarly to the procedure used for fractured thin sheets~\cite{VVV2010,TM2011}. Minimization yields a relation between the optimal $N$ value and the non-dimensional number $\Gamma / (\sigma_0 \et R_i) $
\begin{equation} \label{eq:modelRes}
	\frac{ N \sin^2 (2 \pi /N) }{\sin (2 \pi / N) + 2 \pi  \cos (2 \pi / N)} =  \frac{\Gamma}{\sigma_0 \et R_i} 
\end{equation}

The result of this model is shown with the data in Fig.~\ref{fig:modWL}. The comparison is obtained by adjusting the value of the characteristic length $\ell_c =  \Gamma / (\sigma_0 \et)$. The model shows a decrease of the wavelength of the pattern for small $R_i$ as a result of the geometry: it is a consequence of the polygonal nature of the pattern for small number of cracks. If the number of cracks $N$ is large, Eq.~\ref{eq:modelRes} yields $\Lambda = 2 \pi R_i / N \sim \ell_c$. From our set of data, it is difficult to draw a clear conclusion on the stabilization of $\Lambda$ towards a constant value because the standard deviation from the mean value remains large. We did not perform experiments at larger values of $R_i$ because of the size limitation of the experimental setup.  

A key parameter in the model is the length $\ell_c = \Gamma / ( \sigma_0 \et)$. Here $\et$ is the critical tensile strain when cracks develop and it has been estimated to be of the order of $s_c / d$.  The cohesion stress $\sigma_0$ is associated with the force exerted by each liquid bridge. Macroscopic measurement in tension cell~\cite{PC1997} or in bending~\cite{RB2018} have shown that the scaling $\sigma_0 \sim \gamma / d$ expected from the liquid bridge force~\cite{Rum1970,Mol1975,GTH2003} is effective. The fracture energy is the energy associated with the creation of crack surface. It is obviously related to the energy required to break capillary bridges but other dissipative processes, such as the friction between individual grains as the pile of grains rearrange when crack extends, are also likely to contribute. The energy associated with the breaking of a single bridge is  $U_{LB} = \int F_{LB} ds$ yielding $U_{LB} = \pi \gamma d^{1/2} (2 V_{LB})^{1/2} \mathcal{C}$ where $\mathcal{C}$ is an order 1 slowly varying function of $(V_{LB} / d^3)$. For a fixed liquid content $w$, we obtain $U_{LB} \sim \gamma d^2$ and thus the fracture energy per unit surface $\Gamma$ is expected to be proportional to $\gamma$. The characteristic length $\ell_c = \Gamma / (\sigma_0 \et)$ is then expected to scale like $d$ and thus we expect a pattern that varies with the grain size $d$. This is not what we observe in the experiment. This suggests the need for further exploration of the process of fracture propagation in cohesive granular media. A more detailed exploration of the nature of dissipative processes and of the role of heterogeneity is evidently needed to gain a better understanding of the process of fracture.

\section{Conclusion}

We have shown that a pattern of radial cracks develops when an indenter is pushed quasistatically into a wet granular pile. As the indenter penetrates the granular medium, grains are pushed sideways, and multiple cracks similar to the mode I cracks of solid materials are observed around the indenter. Using the strain field measurements obtained from X-ray microtomography scans, we have shown that the cracks appear in the domain of largest dilation. This suggest a link between the local change of volume of the sample and the apparition of radial cracks. We have discussed the link between the local density and the capillary bridge extension, and shown that the order of magnitude of the mean inter-grain distance inferred from local density at the onset of fracture is consistent with the maximal bridge extension. However, a more detailed imaging would certainly yield additional knowledge on the heterogeneity of the medium, that is likely to play a major role in the nucleation of cracks. We have also characterized some features of the crack pattern. We find that the mean number of radial cracks observed in the pattern increases with the radius of the indenter $R_i$ but does not vary with grain size $d$. This last feature is the most surprising as $d$ is the lengthscale that sets the main material properties. Here also further studies are needed to gain a better insight into the physics of the system. To reduce the dispersion of the result, a more controlled preparation method should be used to obtain a more controlled initial state. Also a more detailed investigation on the crack properties, including a determination of the fracture energy $\Gamma$ is certainly needed. 
As a final word, we emphasize the novelty and difficulties associated with the study of cracks in granular material: the strongly heterogeneous nature of the material and its ability to experience very large deformation (\textit{i.e.} flow) concomitantly with fractures makes it a particularly challenging material to test our understanding of fracture.

\bibliography{BibGrainsPRE}

\begin{thebibliography}{33}%
\makeatletter
\providecommand \@ifxundefined [1]{%
 \@ifx{#1\undefined}
}%
\providecommand \@ifnum [1]{%
 \ifnum #1\expandafter \@firstoftwo
 \else \expandafter \@secondoftwo
 \fi
}%
\providecommand \@ifx [1]{%
 \ifx #1\expandafter \@firstoftwo
 \else \expandafter \@secondoftwo
 \fi
}%
\providecommand \natexlab [1]{#1}%
\providecommand \enquote  [1]{``#1''}%
\providecommand \bibnamefont  [1]{#1}%
\providecommand \bibfnamefont [1]{#1}%
\providecommand \citenamefont [1]{#1}%
\providecommand \href@noop [0]{\@secondoftwo}%
\providecommand \href [0]{\begingroup \@sanitize@url \@href}%
\providecommand \@href[1]{\@@startlink{#1}\@@href}%
\providecommand \@@href[1]{\endgroup#1\@@endlink}%
\providecommand \@sanitize@url [0]{\catcode `\\12\catcode `\$12\catcode `\&12\catcode `\#12\catcode `\^12\catcode `\_12\catcode `\%12\relax}%
\providecommand \@@startlink[1]{}%
\providecommand \@@endlink[0]{}%
\providecommand \url  [0]{\begingroup\@sanitize@url \@url }%
\providecommand \@url [1]{\endgroup\@href {#1}{\urlprefix }}%
\providecommand \urlprefix  [0]{URL }%
\providecommand \Eprint [0]{\href }%
\providecommand \doibase [0]{https://doi.org/}%
\providecommand \selectlanguage [0]{\@gobble}%
\providecommand \bibinfo  [0]{\@secondoftwo}%
\providecommand \bibfield  [0]{\@secondoftwo}%
\providecommand \translation [1]{[#1]}%
\providecommand \BibitemOpen [0]{}%
\providecommand \bibitemStop [0]{}%
\providecommand \bibitemNoStop [0]{.\EOS\space}%
\providecommand \EOS [0]{\spacefactor3000\relax}%
\providecommand \BibitemShut  [1]{\csname bibitem#1\endcsname}%
\let\auto@bib@innerbib\@empty
\bibitem [{\citenamefont {Cottrino}\ \emph {et~al.}(2013)\citenamefont {Cottrino}, \citenamefont {Jorand}, \citenamefont {Maire},\ and\ \citenamefont {Adrien}}]{CJM2013}%
  \BibitemOpen
  \bibfield  {author} {\bibinfo {author} {\bibfnamefont {S.}~\bibnamefont {Cottrino}}, \bibinfo {author} {\bibfnamefont {Y.}~\bibnamefont {Jorand}}, \bibinfo {author} {\bibfnamefont {E.}~\bibnamefont {Maire}},\ and\ \bibinfo {author} {\bibfnamefont {J.}~\bibnamefont {Adrien}},\ }\href@noop {} {\bibfield  {journal} {\bibinfo  {journal} {Materials characterization}\ }\textbf {\bibinfo {volume} {81}},\ \bibinfo {pages} {111} (\bibinfo {year} {2013})}\BibitemShut {NoStop}%
\bibitem [{\citenamefont {Baligh}\ \emph {et~al.}(1980)\citenamefont {Baligh}, \citenamefont {Ladd},\ and\ \citenamefont {Vivatrat}}]{BLV1980}%
  \BibitemOpen
  \bibfield  {author} {\bibinfo {author} {\bibfnamefont {M.~M.}\ \bibnamefont {Baligh}}, \bibinfo {author} {\bibfnamefont {C.~C.}\ \bibnamefont {Ladd}},\ and\ \bibinfo {author} {\bibfnamefont {V.}~\bibnamefont {Vivatrat}},\ }\href@noop {} {\bibfield  {journal} {\bibinfo  {journal} {Journal of the Geotechnical Engineering Division}\ }\textbf {\bibinfo {volume} {106}},\ \bibinfo {pages} {447} (\bibinfo {year} {1980})}\BibitemShut {NoStop}%
\bibitem [{\citenamefont {Mehrabi}\ \emph {et~al.}(2021)\citenamefont {Mehrabi}, \citenamefont {Hassanpour},\ and\ \citenamefont {Bayly}}]{MHB2021}%
  \BibitemOpen
  \bibfield  {author} {\bibinfo {author} {\bibfnamefont {M.}~\bibnamefont {Mehrabi}}, \bibinfo {author} {\bibfnamefont {A.}~\bibnamefont {Hassanpour}},\ and\ \bibinfo {author} {\bibfnamefont {A.}~\bibnamefont {Bayly}},\ }\href@noop {} {\bibfield  {journal} {\bibinfo  {journal} {Powder Technology}\ }\textbf {\bibinfo {volume} {385}},\ \bibinfo {pages} {250} (\bibinfo {year} {2021})}\BibitemShut {NoStop}%
\bibitem [{\citenamefont {Sharpe}\ \emph {et~al.}(2015)\citenamefont {Sharpe}, \citenamefont {Kuckuk},\ and\ \citenamefont {Goldman}}]{SKG2015}%
  \BibitemOpen
  \bibfield  {author} {\bibinfo {author} {\bibfnamefont {S.~S.}\ \bibnamefont {Sharpe}}, \bibinfo {author} {\bibfnamefont {R.}~\bibnamefont {Kuckuk}},\ and\ \bibinfo {author} {\bibfnamefont {D.~I.}\ \bibnamefont {Goldman}},\ }\href@noop {} {\bibfield  {journal} {\bibinfo  {journal} {Physical Biology}\ }\textbf {\bibinfo {volume} {12}},\ \bibinfo {pages} {046009} (\bibinfo {year} {2015})}\BibitemShut {NoStop}%
\bibitem [{\citenamefont {Katsuragi}(2016)}]{Katsuragi2016}%
  \BibitemOpen
  \bibfield  {author} {\bibinfo {author} {\bibfnamefont {H.}~\bibnamefont {Katsuragi}},\ }\href@noop {} {\emph {\bibinfo {title} {Physics of Soft Impact and Cratering}}}\ (\bibinfo  {publisher} {Springer},\ \bibinfo {address} {Tokyo},\ \bibinfo {year} {2016})\BibitemShut {NoStop}%
\bibitem [{\citenamefont {Kang}\ \emph {et~al.}(2018)\citenamefont {Kang}, \citenamefont {Feng}, \citenamefont {Liu},\ and\ \citenamefont {Blumenfeld}}]{KFL2018}%
  \BibitemOpen
  \bibfield  {author} {\bibinfo {author} {\bibfnamefont {W.}~\bibnamefont {Kang}}, \bibinfo {author} {\bibfnamefont {Y.}~\bibnamefont {Feng}}, \bibinfo {author} {\bibfnamefont {C.}~\bibnamefont {Liu}},\ and\ \bibinfo {author} {\bibfnamefont {R.}~\bibnamefont {Blumenfeld}},\ }\href@noop {} {\bibfield  {journal} {\bibinfo  {journal} {Nature communications}\ }\textbf {\bibinfo {volume} {9}},\ \bibinfo {pages} {1} (\bibinfo {year} {2018})}\BibitemShut {NoStop}%
\bibitem [{\citenamefont {McDonald}\ \emph {et~al.}(2006)\citenamefont {McDonald}, \citenamefont {Schneider}, \citenamefont {Cocks},\ and\ \citenamefont {Withers}}]{MSC2006}%
  \BibitemOpen
  \bibfield  {author} {\bibinfo {author} {\bibfnamefont {S.~A.}\ \bibnamefont {McDonald}}, \bibinfo {author} {\bibfnamefont {L.~C.~R.}\ \bibnamefont {Schneider}}, \bibinfo {author} {\bibfnamefont {A.~C.~F.}\ \bibnamefont {Cocks}},\ and\ \bibinfo {author} {\bibfnamefont {P.~J.}\ \bibnamefont {Withers}},\ }\href@noop {} {\bibfield  {journal} {\bibinfo  {journal} {Scripta Materialia}\ }\textbf {\bibinfo {volume} {54}},\ \bibinfo {pages} {191} (\bibinfo {year} {2006})}\BibitemShut {NoStop}%
\bibitem [{\citenamefont {Mitarai}\ and\ \citenamefont {Nori}(2006)}]{MN2006}%
  \BibitemOpen
  \bibfield  {author} {\bibinfo {author} {\bibfnamefont {N.}~\bibnamefont {Mitarai}}\ and\ \bibinfo {author} {\bibfnamefont {F.}~\bibnamefont {Nori}},\ }\href@noop {} {\bibfield  {journal} {\bibinfo  {journal} {Advances in Physics}\ }\textbf {\bibinfo {volume} {55}},\ \bibinfo {pages} {1} (\bibinfo {year} {2006})}\BibitemShut {NoStop}%
\bibitem [{\citenamefont {Herminghaus}(2005)}]{Her2005}%
  \BibitemOpen
  \bibfield  {author} {\bibinfo {author} {\bibfnamefont {S.}~\bibnamefont {Herminghaus}},\ }\href@noop {} {\bibfield  {journal} {\bibinfo  {journal} {Advances in Physics}\ }\textbf {\bibinfo {volume} {54}},\ \bibinfo {pages} {221} (\bibinfo {year} {2005})}\BibitemShut {NoStop}%
\bibitem [{\citenamefont {Richefeu}\ \emph {et~al.}(2006)\citenamefont {Richefeu}, \citenamefont {El~Youssoufi},\ and\ \citenamefont {Radja\"{\i}}}]{RER2006}%
  \BibitemOpen
  \bibfield  {author} {\bibinfo {author} {\bibfnamefont {V.}~\bibnamefont {Richefeu}}, \bibinfo {author} {\bibfnamefont {M.~S.}\ \bibnamefont {El~Youssoufi}},\ and\ \bibinfo {author} {\bibfnamefont {F.}~\bibnamefont {Radja\"{\i}}},\ }\href {https://doi.org/10.1103/PhysRevE.73.051304} {\bibfield  {journal} {\bibinfo  {journal} {Phys. Rev. E}\ }\textbf {\bibinfo {volume} {73}},\ \bibinfo {pages} {051304} (\bibinfo {year} {2006})}\BibitemShut {NoStop}%
\bibitem [{\citenamefont {Divoux}\ \emph {et~al.}(2016)\citenamefont {Divoux}, \citenamefont {Fardin}, \citenamefont {Manneville},\ and\ \citenamefont {Lerouge}}]{DFM2016}%
  \BibitemOpen
  \bibfield  {author} {\bibinfo {author} {\bibfnamefont {T.}~\bibnamefont {Divoux}}, \bibinfo {author} {\bibfnamefont {M.~A.}\ \bibnamefont {Fardin}}, \bibinfo {author} {\bibfnamefont {S.}~\bibnamefont {Manneville}},\ and\ \bibinfo {author} {\bibfnamefont {S.}~\bibnamefont {Lerouge}},\ }\href@noop {} {\bibfield  {journal} {\bibinfo  {journal} {Annual Review of Fluid Mechanics}\ }\textbf {\bibinfo {volume} {48}},\ \bibinfo {pages} {81} (\bibinfo {year} {2016})}\BibitemShut {NoStop}%
\bibitem [{\citenamefont {Mandal}\ \emph {et~al.}(2021)\citenamefont {Mandal}, \citenamefont {Nicolas},\ and\ \citenamefont {Pouliquen}}]{MNP2021}%
  \BibitemOpen
  \bibfield  {author} {\bibinfo {author} {\bibfnamefont {S.}~\bibnamefont {Mandal}}, \bibinfo {author} {\bibfnamefont {M.}~\bibnamefont {Nicolas}},\ and\ \bibinfo {author} {\bibfnamefont {O.}~\bibnamefont {Pouliquen}},\ }\href {https://doi.org/10.1103/PhysRevX.11.021017} {\bibfield  {journal} {\bibinfo  {journal} {Phys. Rev. X}\ }\textbf {\bibinfo {volume} {11}},\ \bibinfo {pages} {021017} (\bibinfo {year} {2021})}\BibitemShut {NoStop}%
\bibitem [{\citenamefont {Broberg}(1999)}]{Broberg1999}%
  \BibitemOpen
  \bibfield  {author} {\bibinfo {author} {\bibfnamefont {K.~B.}\ \bibnamefont {Broberg}},\ }\href@noop {} {\emph {\bibinfo {title} {{Cracks and fracture}}}}\ (\bibinfo  {publisher} {Academic Press},\ \bibinfo {address} {San Diego},\ \bibinfo {year} {1999})\BibitemShut {NoStop}%
\bibitem [{\citenamefont {Pierrat}\ and\ \citenamefont {Caram}(1997)}]{PC1997}%
  \BibitemOpen
  \bibfield  {author} {\bibinfo {author} {\bibfnamefont {P.}~\bibnamefont {Pierrat}}\ and\ \bibinfo {author} {\bibfnamefont {H.~S.}\ \bibnamefont {Caram}},\ }\href@noop {} {\bibfield  {journal} {\bibinfo  {journal} {Powder Technology}\ }\textbf {\bibinfo {volume} {91}},\ \bibinfo {pages} {83} (\bibinfo {year} {1997})}\BibitemShut {NoStop}%
\bibitem [{\citenamefont {Scheel}\ \emph {et~al.}(2008)\citenamefont {Scheel}, \citenamefont {Seemann}, \citenamefont {Brinkmann}, \citenamefont {Di~Michiel}, \citenamefont {Sheppard}, \citenamefont {Breidenbach},\ and\ \citenamefont {Herminghaus}}]{SSB2008}%
  \BibitemOpen
  \bibfield  {author} {\bibinfo {author} {\bibfnamefont {M.}~\bibnamefont {Scheel}}, \bibinfo {author} {\bibfnamefont {R.}~\bibnamefont {Seemann}}, \bibinfo {author} {\bibfnamefont {M.}~\bibnamefont {Brinkmann}}, \bibinfo {author} {\bibfnamefont {M.}~\bibnamefont {Di~Michiel}}, \bibinfo {author} {\bibfnamefont {A.}~\bibnamefont {Sheppard}}, \bibinfo {author} {\bibfnamefont {B.}~\bibnamefont {Breidenbach}},\ and\ \bibinfo {author} {\bibfnamefont {S.}~\bibnamefont {Herminghaus}},\ }\href@noop {} {\bibfield  {journal} {\bibinfo  {journal} {Nature materials}\ }\textbf {\bibinfo {volume} {7}},\ \bibinfo {pages} {189} (\bibinfo {year} {2008})}\BibitemShut {NoStop}%
\bibitem [{\citenamefont {Raux}\ and\ \citenamefont {Biance}(2018)}]{RB2018}%
  \BibitemOpen
  \bibfield  {author} {\bibinfo {author} {\bibfnamefont {P.~S.}\ \bibnamefont {Raux}}\ and\ \bibinfo {author} {\bibfnamefont {A.-L.}\ \bibnamefont {Biance}},\ }\href@noop {} {\bibfield  {journal} {\bibinfo  {journal} {Physical Review Fluids}\ }\textbf {\bibinfo {volume} {3}},\ \bibinfo {pages} {014301} (\bibinfo {year} {2018})}\BibitemShut {NoStop}%
\bibitem [{\citenamefont {Roch{\'e}}\ \emph {et~al.}(2013)\citenamefont {Roch{\'e}}, \citenamefont {Myftiu}, \citenamefont {Johnston}, \citenamefont {Kim},\ and\ \citenamefont {Stone}}]{RMJ2013}%
  \BibitemOpen
  \bibfield  {author} {\bibinfo {author} {\bibfnamefont {M.}~\bibnamefont {Roch{\'e}}}, \bibinfo {author} {\bibfnamefont {E.}~\bibnamefont {Myftiu}}, \bibinfo {author} {\bibfnamefont {M.~C.}\ \bibnamefont {Johnston}}, \bibinfo {author} {\bibfnamefont {P.}~\bibnamefont {Kim}},\ and\ \bibinfo {author} {\bibfnamefont {H.~A.}\ \bibnamefont {Stone}},\ }\href@noop {} {\bibfield  {journal} {\bibinfo  {journal} {Physical Review Letters}\ }\textbf {\bibinfo {volume} {110}},\ \bibinfo {pages} {148304} (\bibinfo {year} {2013})}\BibitemShut {NoStop}%
\bibitem [{\citenamefont {Ball}\ \emph {et~al.}(2022)\citenamefont {Ball}, \citenamefont {Balmforth}, \citenamefont {Dufresne},\ and\ \citenamefont {Morris}}]{BBD2022}%
  \BibitemOpen
  \bibfield  {author} {\bibinfo {author} {\bibfnamefont {T.~V.}\ \bibnamefont {Ball}}, \bibinfo {author} {\bibfnamefont {N.~J.}\ \bibnamefont {Balmforth}}, \bibinfo {author} {\bibfnamefont {A.~P.}\ \bibnamefont {Dufresne}},\ and\ \bibinfo {author} {\bibfnamefont {S.~W.}\ \bibnamefont {Morris}},\ }\href {https://doi.org/10.1017/jfm.2021.961} {\bibfield  {journal} {\bibinfo  {journal} {Journal of Fluid Mechanics}\ }\textbf {\bibinfo {volume} {934}},\ \bibinfo {pages} {A31} (\bibinfo {year} {2022})}\BibitemShut {NoStop}%
\bibitem [{\citenamefont {Sharma}\ \emph {et~al.}(2024)\citenamefont {Sharma}, \citenamefont {Sarlin}, \citenamefont {Xing}, \citenamefont {Morize}, \citenamefont {Gondret},\ and\ \citenamefont {Sauret}}]{SSX2024}%
  \BibitemOpen
  \bibfield  {author} {\bibinfo {author} {\bibfnamefont {R.~S.}\ \bibnamefont {Sharma}}, \bibinfo {author} {\bibfnamefont {W.}~\bibnamefont {Sarlin}}, \bibinfo {author} {\bibfnamefont {L.}~\bibnamefont {Xing}}, \bibinfo {author} {\bibfnamefont {C.}~\bibnamefont {Morize}}, \bibinfo {author} {\bibfnamefont {P.}~\bibnamefont {Gondret}},\ and\ \bibinfo {author} {\bibfnamefont {A.}~\bibnamefont {Sauret}},\ }\href@noop {} {\bibfield  {journal} {\bibinfo  {journal} {Physical Review Fluids}\ }\textbf {\bibinfo {volume} {9}},\ \bibinfo {pages} {074301} (\bibinfo {year} {2024})}\BibitemShut {NoStop}%
\bibitem [{\citenamefont {Costes}\ \emph {et~al.}(1970)\citenamefont {Costes}, \citenamefont {Carrier}, \citenamefont {Mitchell},\ and\ \citenamefont {Scott}}]{CCM1970}%
  \BibitemOpen
  \bibfield  {author} {\bibinfo {author} {\bibfnamefont {N.}~\bibnamefont {Costes}}, \bibinfo {author} {\bibfnamefont {W.}~\bibnamefont {Carrier}}, \bibinfo {author} {\bibfnamefont {J.}~\bibnamefont {Mitchell}},\ and\ \bibinfo {author} {\bibfnamefont {R.}~\bibnamefont {Scott}},\ }\href@noop {} {\bibfield  {journal} {\bibinfo  {journal} {Science}\ }\textbf {\bibinfo {volume} {167}},\ \bibinfo {pages} {739} (\bibinfo {year} {1970})}\BibitemShut {NoStop}%
\bibitem [{\citenamefont {Lin}\ \emph {et~al.}(2009)\citenamefont {Lin}, \citenamefont {Otim}, \citenamefont {Lenhart}, \citenamefont {Cole},\ and\ \citenamefont {Shull}}]{LOL2009}%
  \BibitemOpen
  \bibfield  {author} {\bibinfo {author} {\bibfnamefont {W.-C.}\ \bibnamefont {Lin}}, \bibinfo {author} {\bibfnamefont {K.~J.}\ \bibnamefont {Otim}}, \bibinfo {author} {\bibfnamefont {J.~L.}\ \bibnamefont {Lenhart}}, \bibinfo {author} {\bibfnamefont {P.~J.}\ \bibnamefont {Cole}},\ and\ \bibinfo {author} {\bibfnamefont {K.~R.}\ \bibnamefont {Shull}},\ }\href@noop {} {\bibfield  {journal} {\bibinfo  {journal} {Journal of Materials Research}\ }\textbf {\bibinfo {volume} {24}},\ \bibinfo {pages} {957} (\bibinfo {year} {2009})}\BibitemShut {NoStop}%
\bibitem [{\citenamefont {Rouxel}\ \emph {et~al.}(2021)\citenamefont {Rouxel}, \citenamefont {Jang},\ and\ \citenamefont {Ramamurty}}]{RJR2021}%
  \BibitemOpen
  \bibfield  {author} {\bibinfo {author} {\bibfnamefont {T.}~\bibnamefont {Rouxel}}, \bibinfo {author} {\bibfnamefont {J.-i.}\ \bibnamefont {Jang}},\ and\ \bibinfo {author} {\bibfnamefont {U.}~\bibnamefont {Ramamurty}},\ }\href@noop {} {\bibfield  {journal} {\bibinfo  {journal} {Progress in Materials Science}\ }\textbf {\bibinfo {volume} {121}},\ \bibinfo {pages} {100834} (\bibinfo {year} {2021})}\BibitemShut {NoStop}%
\bibitem [{\citenamefont {Feng}\ and\ \citenamefont {Yu}(1998)}]{FY1998}%
  \BibitemOpen
  \bibfield  {author} {\bibinfo {author} {\bibfnamefont {C.~L.}\ \bibnamefont {Feng}}\ and\ \bibinfo {author} {\bibfnamefont {A.~D.}\ \bibnamefont {Yu}},\ }\href@noop {} {\bibfield  {journal} {\bibinfo  {journal} {Powder Technology}\ }\textbf {\bibinfo {volume} {99}},\ \bibinfo {pages} {22} (\bibinfo {year} {1998})}\BibitemShut {NoStop}%
\bibitem [{\citenamefont {Fournier}\ \emph {et~al.}(2005)\citenamefont {Fournier}, \citenamefont {Geromichalos}, \citenamefont {Herminghaus}, \citenamefont {Kohonen}, \citenamefont {Mugele}, \citenamefont {Scheel}, \citenamefont {Schulz}, \citenamefont {Schulz}, \citenamefont {Schier}, \citenamefont {Seemann} \emph {et~al.}}]{FGH2005}%
  \BibitemOpen
  \bibfield  {author} {\bibinfo {author} {\bibfnamefont {Z.}~\bibnamefont {Fournier}}, \bibinfo {author} {\bibfnamefont {D.}~\bibnamefont {Geromichalos}}, \bibinfo {author} {\bibfnamefont {S.}~\bibnamefont {Herminghaus}}, \bibinfo {author} {\bibfnamefont {M.}~\bibnamefont {Kohonen}}, \bibinfo {author} {\bibfnamefont {F.}~\bibnamefont {Mugele}}, \bibinfo {author} {\bibfnamefont {M.}~\bibnamefont {Scheel}}, \bibinfo {author} {\bibfnamefont {M.}~\bibnamefont {Schulz}}, \bibinfo {author} {\bibfnamefont {B.}~\bibnamefont {Schulz}}, \bibinfo {author} {\bibfnamefont {C.}~\bibnamefont {Schier}}, \bibinfo {author} {\bibfnamefont {R.}~\bibnamefont {Seemann}}, \emph {et~al.},\ }\href {https://doi.org/10.1088/0953-8984/17/9/013} {\bibfield  {journal} {\bibinfo  {journal} {Journal of Physics: Condensed Matter}\ }\textbf {\bibinfo {volume} {17}},\ \bibinfo {pages} {S477} (\bibinfo {year} {2005})}\BibitemShut {NoStop}%
\bibitem [{\citenamefont {Holzapfel}(2000)}]{Holzapfel2000}%
  \BibitemOpen
  \bibfield  {author} {\bibinfo {author} {\bibfnamefont {G.~A.}\ \bibnamefont {Holzapfel}},\ }\href@noop {} {\emph {\bibinfo {title} {Nonlinear solid mechanics}}}\ (\bibinfo  {publisher} {Wiley},\ \bibinfo {address} {Chichester, England},\ \bibinfo {year} {2000})\BibitemShut {NoStop}%
\bibitem [{\citenamefont {Willett}\ \emph {et~al.}(2000)\citenamefont {Willett}, \citenamefont {Adams}, \citenamefont {Johnson},\ and\ \citenamefont {Seville}}]{WAJ2000}%
  \BibitemOpen
  \bibfield  {author} {\bibinfo {author} {\bibfnamefont {C.~D.}\ \bibnamefont {Willett}}, \bibinfo {author} {\bibfnamefont {M.~J.}\ \bibnamefont {Adams}}, \bibinfo {author} {\bibfnamefont {S.~A.}\ \bibnamefont {Johnson}},\ and\ \bibinfo {author} {\bibfnamefont {J.~P.~K.}\ \bibnamefont {Seville}},\ }\href@noop {} {\bibfield  {journal} {\bibinfo  {journal} {Langmuir}\ }\textbf {\bibinfo {volume} {16}},\ \bibinfo {pages} {9396} (\bibinfo {year} {2000})}\BibitemShut {NoStop}%
\bibitem [{\citenamefont {Bahr}\ \emph {et~al.}(2010)\citenamefont {Bahr}, \citenamefont {Weiss}, \citenamefont {Bahr}, \citenamefont {Hofmann}, \citenamefont {Fischer}, \citenamefont {Lampenscherf},\ and\ \citenamefont {Balke}}]{BWB2010}%
  \BibitemOpen
  \bibfield  {author} {\bibinfo {author} {\bibfnamefont {H.-A.}\ \bibnamefont {Bahr}}, \bibinfo {author} {\bibfnamefont {H.-J.}\ \bibnamefont {Weiss}}, \bibinfo {author} {\bibfnamefont {U.}~\bibnamefont {Bahr}}, \bibinfo {author} {\bibfnamefont {M.}~\bibnamefont {Hofmann}}, \bibinfo {author} {\bibfnamefont {G.}~\bibnamefont {Fischer}}, \bibinfo {author} {\bibfnamefont {S.}~\bibnamefont {Lampenscherf}},\ and\ \bibinfo {author} {\bibfnamefont {H.}~\bibnamefont {Balke}},\ }\href@noop {} {\bibfield  {journal} {\bibinfo  {journal} {Journal of the Mechanics and Physics of Solids}\ }\textbf {\bibinfo {volume} {58}},\ \bibinfo {pages} {1411} (\bibinfo {year} {2010})}\BibitemShut {NoStop}%
\bibitem [{\citenamefont {Vermorel}\ \emph {et~al.}(2010)\citenamefont {Vermorel}, \citenamefont {Vandenberghe},\ and\ \citenamefont {Villermaux}}]{VVV2010}%
  \BibitemOpen
  \bibfield  {author} {\bibinfo {author} {\bibfnamefont {R.}~\bibnamefont {Vermorel}}, \bibinfo {author} {\bibfnamefont {N.}~\bibnamefont {Vandenberghe}},\ and\ \bibinfo {author} {\bibfnamefont {E.}~\bibnamefont {Villermaux}},\ }\href@noop {} {\bibfield  {journal} {\bibinfo  {journal} {Physical Review Letters}\ }\textbf {\bibinfo {volume} {104}},\ \bibinfo {pages} {175502} (\bibinfo {year} {2010})}\BibitemShut {NoStop}%
\bibitem [{\citenamefont {Vandenberghe}\ \emph {et~al.}(2013)\citenamefont {Vandenberghe}, \citenamefont {Vermorel},\ and\ \citenamefont {Villermaux}}]{VVV2013}%
  \BibitemOpen
  \bibfield  {author} {\bibinfo {author} {\bibfnamefont {N.}~\bibnamefont {Vandenberghe}}, \bibinfo {author} {\bibfnamefont {R.}~\bibnamefont {Vermorel}},\ and\ \bibinfo {author} {\bibfnamefont {E.}~\bibnamefont {Villermaux}},\ }\href@noop {} {\bibfield  {journal} {\bibinfo  {journal} {Physical Review Letters}\ }\textbf {\bibinfo {volume} {110}},\ \bibinfo {pages} {174302} (\bibinfo {year} {2013})}\BibitemShut {NoStop}%
\bibitem [{\citenamefont {Tallinen}\ and\ \citenamefont {Mahadevan}(2011)}]{TM2011}%
  \BibitemOpen
  \bibfield  {author} {\bibinfo {author} {\bibfnamefont {T.}~\bibnamefont {Tallinen}}\ and\ \bibinfo {author} {\bibfnamefont {L.}~\bibnamefont {Mahadevan}},\ }\href@noop {} {\bibfield  {journal} {\bibinfo  {journal} {Physical Review Letters}\ }\textbf {\bibinfo {volume} {107}},\ \bibinfo {pages} {245502} (\bibinfo {year} {2011})}\BibitemShut {NoStop}%
\bibitem [{\citenamefont {Rumpf}(1970)}]{Rum1970}%
  \BibitemOpen
  \bibfield  {author} {\bibinfo {author} {\bibfnamefont {H.}~\bibnamefont {Rumpf}},\ }\href {https://doi.org/10.1002/cite.330420806} {\bibfield  {journal} {\bibinfo  {journal} {Chemie Ingenieur Technik}\ }\textbf {\bibinfo {volume} {42}},\ \bibinfo {pages} {538} (\bibinfo {year} {1970})}\BibitemShut {NoStop}%
\bibitem [{\citenamefont {Molerus}(1975)}]{Mol1975}%
  \BibitemOpen
  \bibfield  {author} {\bibinfo {author} {\bibfnamefont {O.}~\bibnamefont {Molerus}},\ }\href@noop {} {\bibfield  {journal} {\bibinfo  {journal} {Powder Technology}\ }\textbf {\bibinfo {volume} {12}},\ \bibinfo {pages} {259} (\bibinfo {year} {1975})}\BibitemShut {NoStop}%
\bibitem [{\citenamefont {Gr{\"o}ger}\ \emph {et~al.}(2003)\citenamefont {Gr{\"o}ger}, \citenamefont {T{\"u}z{\"u}n},\ and\ \citenamefont {Heyes}}]{GTH2003}%
  \BibitemOpen
  \bibfield  {author} {\bibinfo {author} {\bibfnamefont {T.}~\bibnamefont {Gr{\"o}ger}}, \bibinfo {author} {\bibfnamefont {U.}~\bibnamefont {T{\"u}z{\"u}n}},\ and\ \bibinfo {author} {\bibfnamefont {D.~M.}\ \bibnamefont {Heyes}},\ }\href {https://doi.org/10.1016/S0032-5910(03)00093-7} {\bibfield  {journal} {\bibinfo  {journal} {Powder Technology}\ }\textbf {\bibinfo {volume} {133}},\ \bibinfo {pages} {203} (\bibinfo {year} {2003})}\BibitemShut {NoStop}%
\end{thebibliography}%

\end{document}


\doublespacing

\renewcommand{\figurename}{SI Figure}

\title{Supplementary materials for the paper: the crack pattern of indented granular media}
\maketitle

\section{Additional informations about the material used}

Table \ref{tab:beads} lists the properties of the beads used in the experiments. After being carefully cleaned the spheres are mixed with silicon oil (see main text). 
\begin{table}[ht!]
\begin{center}
\begin{tabular}{|c|c|c|c|c|}
\hline
Reference &  Diameter $d$ ($\mu$m) & Bond number $\mathrm{Bo} = \rho_b g d^2/\gamma$\\
\hline
Silibeads 5211 &  $55\pm15$ & $3.7\times 10^{-3}$  \\
\hline
Silibeads 5212 & $90\pm20$ & $9.9\times 10^{-3}$\\
\hline
Silibeads 5215 &$200\pm50$ & $4.9\times 10^{-2}$\\
\hline
Silibeads 5220 & $358\pm43$ & $0.16$\\
\hline
Silibeads 5218 & $500\pm100$ & $0.31$\\
\hline
Silibeads 201-0470 & $1150\pm150$& $1.6$\\
\hline
\end{tabular}
\caption{Glass bead characteristics.}
\end{center}
\label{tab:beads}
\end{table}

\begin{table}[ht!]
    \centering
    \begin{tabular}{|c|c|c|}
 \hline
 $w$ & $\rho_p$ (kg~m$^{-3}$) & $\phi_0$   \\
 \hline
0.5 & $1450\pm 50$ & $0.58\pm0.02$       \\
 \hline
8 & $1425\pm 50$ & $0.57\pm0.02$       \\
\hline
    \end{tabular}
    \caption{Pile density and packing fraction for the 2 different materials. The interval corresponds to the maximal and minimal values over all the experiments, and is larger than the measurement uncertainty. }
    \label{tab:density}
\end{table}

\section{Crack pattern}

\subsection{Crack patterns}

\begin{figure}[h]
   \centering
    \includegraphics[width=15cm]{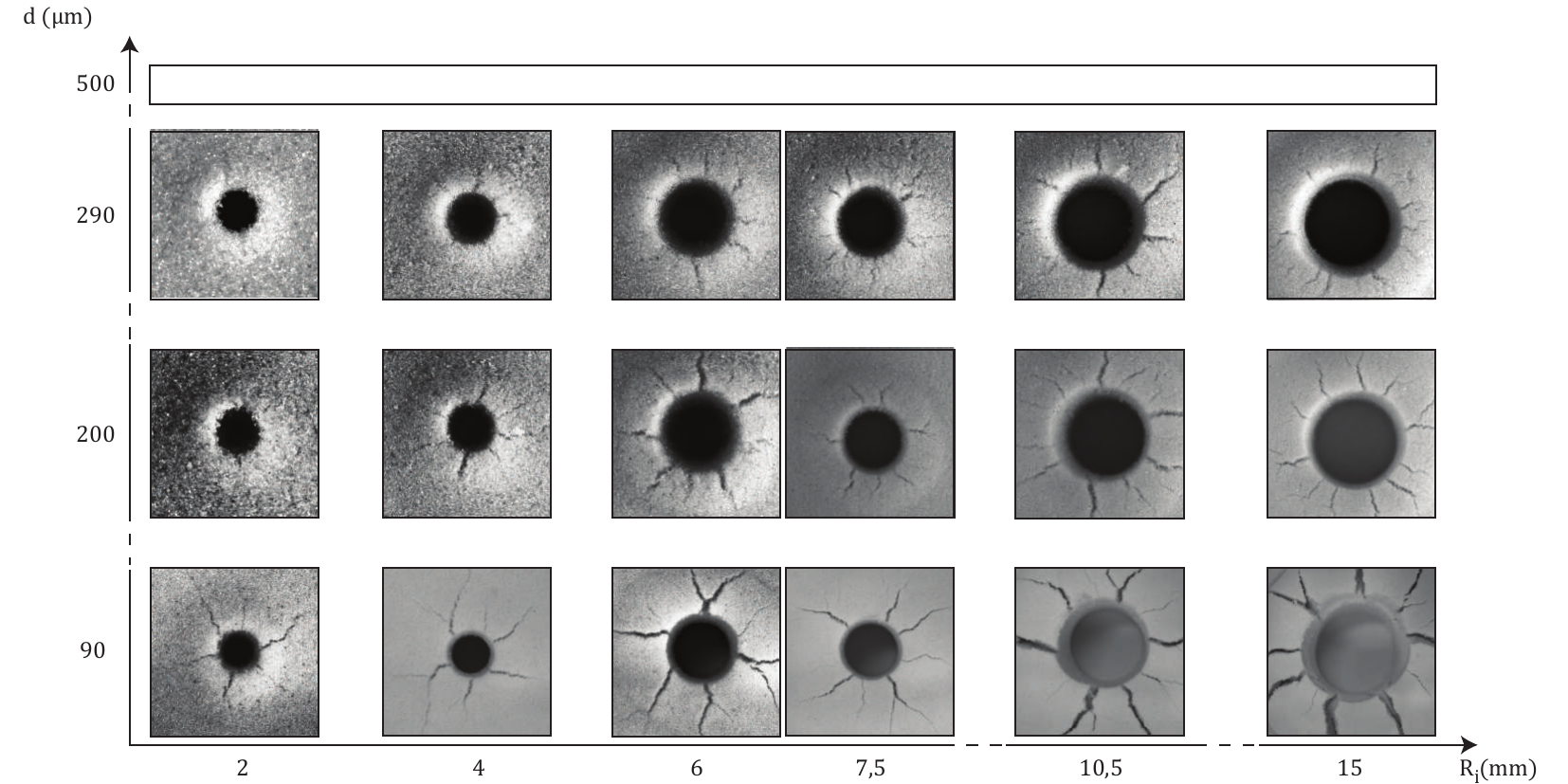}
    \includegraphics[width=15cm]{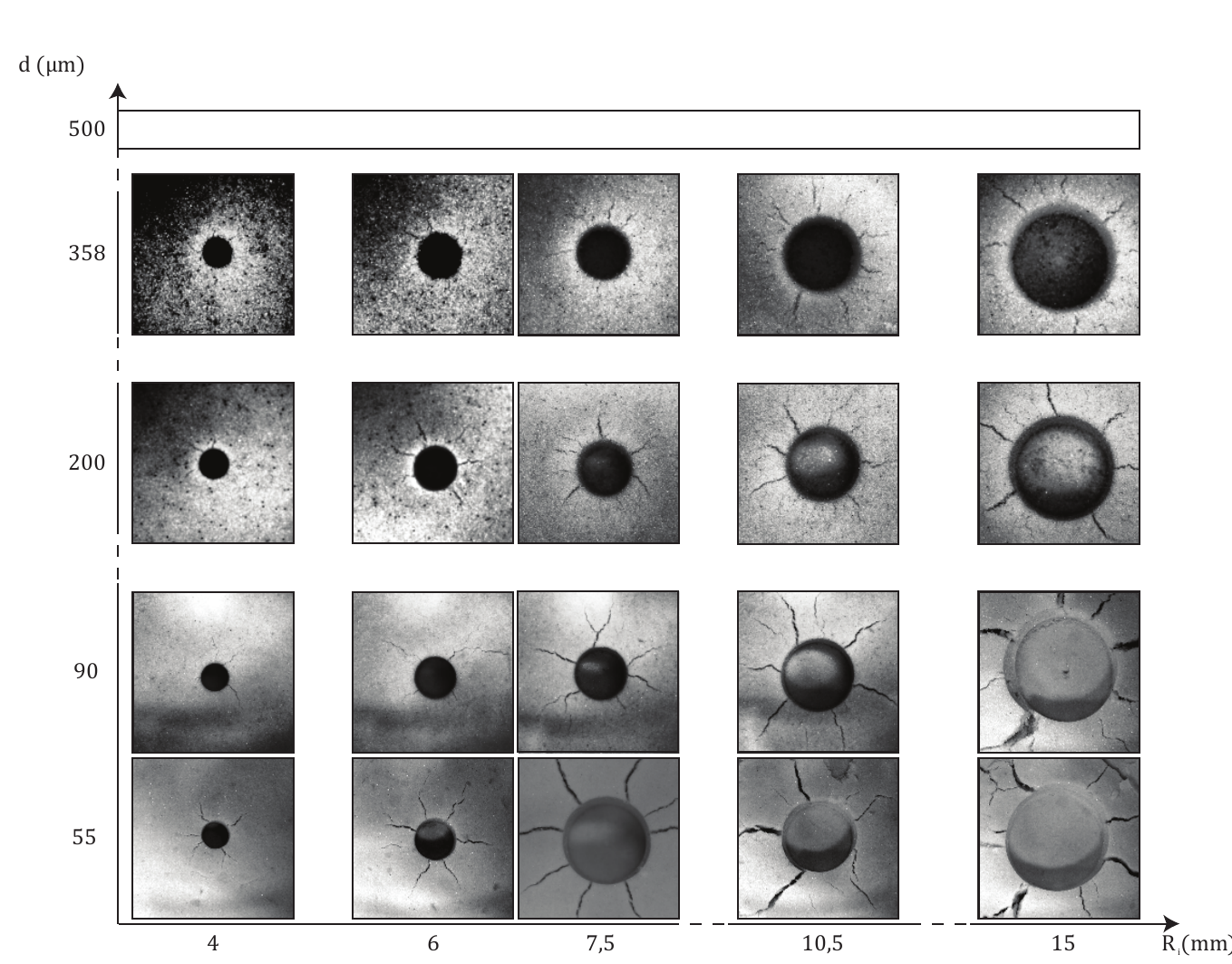}
    \caption{Crack pattern for $w=0.5\%$ (top) and $w=8\%$ (bottom). The scales differ depending on the image (see indenter radius for scale).}
    \label{fig:crackpattern}
\end{figure}

To measure the number of cracks in the indentation experiments, we rely on photographs (SI figure~\ref{fig:crackpattern}). At the end of an experiment, we remove the indenter by carefully lifting it, which does not change the crack pattern. We then take a picture from above, and count all the cracks. In some cases we change the contrast in order to count even the faintest cracks.  

\newpage

\newpage

\section{Displacement and dilation fields}

Displacement fields within samples were measured by tracking the motion of metallic markers dispersed in the sample (see main text for additional details). We observe a radial symmetry, allowing us to compute average displacement over the ortho-radial angle $\theta$ : $u_r$ and $u_z$. These fields are computed in the Lagrangian coordinates $(R,Z)$, corresponding to the initial position of the marker in the pile. We can convert to the current position  $(r,z)$ : $r=R+u_r$, $z=Z+u_z$.

\subsection{Fitting functions}

The experimental values for the displacement exhibit some noise. To compute accurately the derivatives in order to compute strain and dilation field, we fit the experimental curves using ad-hoc functions, and use these functions to compute the derivative. The fitting functions are 
\begin{equation}
    u_{r} (R, Z,z_i) = \left[ a_R(R,z_i) + b_R(R,z_i) Z \right] e^{ c_R(R) ( Z - z_i)^2 },
    \label{eq:fit_ur}
\end{equation}
\begin{equation}
    u_{z} (R, Z,z_i)=\frac{ a_Z(R,z_i) Z^2}{ Z^2 + b_Z(R) Z + c_Z(R)} - a_Z(R,z_i),
    \label{eq:fit_uz}
\end{equation}

with $z_i$ the indenter depth. These functions were chosen so they can fit the displacements at any given point. To obtain them, we fix the radius $R$ and fit the curves $u_{R}(Z)$ and $u_Z(Z)$, getting the values of $a_R$, $b_R$, $c_R$, $a_Z$, $b_Z$ and $c_Z$. These values depend on $R$ (and $z_i$ in some cases) arbitrarily.

\subsection{Dilation field}
We then smooth the fields in the $Z$ direction in order to obtain the derivative by finite difference. We compute the Jacobian of the displacement field, assuming no displacement in the ortho-radial direction $\theta$, and a radial symmetry: 

\begin{equation}
    \nabla_X \underline{u}= 
    \begin{pmatrix}
\frac{\partial u_{r}}{\partial R}+1& 0 & \frac{\partial u_{r}}{\partial Z}\\
0  & \frac{u_{r}}{R} +1&    0\\
\frac{\partial u_z}{\partial R} & 0 & \frac{\partial u_z}{\partial Z}+1
\end{pmatrix}.
\end{equation}

We can then compute the normalized change of volume  $ \Delta V / V_0 = J - 1$, with  $J = \mathrm{det} \,\mathbb{F}$ and  $\mathbb{F} = \mathbb{I} + \nabla_X \underline{u}$. 

\begin{figure*}[ht!]
    \centering
    \includegraphics[width=\linewidth]{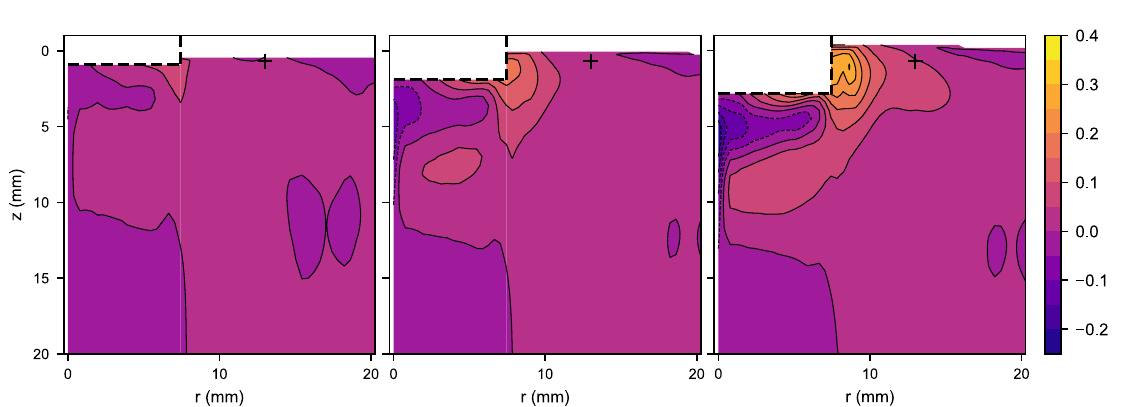}
    \caption{Change in volume $\frac{\Delta V}{V_0}$ for $R_i=7.5$~mm, $d=200~\mu$m, $w=8\%$ and different indentation depth:   $z_i=0.9$~mm (left), $z_i=1.9$~mm (middle) and $z_i=z_c=2.8$~mm (right). The average location of the crack tip at nucleation is shown by the black cross.}
    \label{fig:dilatation_diffzi}
\end{figure*}

In SI figure~\ref{fig:dilatation_diffzi}, we show the change in volume for different $z_i$. We see that the shape of the field stays similar, but the amplitude increases.

